\newif\ifanonymized
\anonymizedfalse

\documentclass[conference]{IEEEtran}
\IEEEoverridecommandlockouts

\usepackage{cite}
\usepackage{amsmath,amssymb,amsfonts}
\usepackage{algorithm}
\usepackage{graphicx}
\usepackage{textcomp}
\usepackage{url}
\usepackage{xcolor}
\usepackage[most]{tcolorbox}
\usepackage{multirow}
\usepackage{stfloats}
\usepackage{tikz}
\usepackage{tabularx}
\usepackage{makecell}
\usepackage{booktabs}
\usepackage{enumitem}
\usepackage[hidelinks]{hyperref}

\newtcolorbox{mybox}[1][]{%
  colback=gray!10,
  colframe=black!50,
  boxrule=0.1pt,
  arc=1pt,
  lefttitle=0.15cm,
  righttitle=0.15cm,
  leftupper=0.15cm,
  rightupper=0.15cm,
  title=Insights from \ifanonymized HotelSoftwareCompany \else  CASBLANCA  \fi,
  breakable,
  #1 
}

\def\BibTeX{{\rm B\kern-.05em{\sc i\kern-.025em b}\kern-.08em
    T\kern-.1667em\lower.7ex\hbox{E}\kern-.125emX}}
\begin{document}

\title{RAGnaroX: A Secure, Local-Hosted ChatOps Assistant Using Small Language Models\\
\thanks{This work was carried out within the \ifanonymized XXXX \else ITEA4 GENIUS \fi project, funded by \ifanonymized XXXX \else FFG (grant 921318). \fi}
}

\author{
\ifanonymized
  \IEEEauthorblockN{Anonymous Author(s)}
  \IEEEauthorblockA{Affiliation(s) withheld for peer review}
\else
\IEEEauthorblockN{Benedikt Dornauer}
\IEEEauthorblockA{
    University of Innsbruck\\
    Innsbruck, Austria\\
    Email: benedikt.dornauer@uibk.ac.at\\
    ORCID: 0000-0002-7713-4686
}
\and
\IEEEauthorblockN{Mircea-Cristian Racasan}
\IEEEauthorblockA{
    c.c.com Moser GmbH\\
    Teslastraße 4, 8074 Grambach, Austria\\
    Email: mracasan@cccom.at\\
    ORCID: 0009-0008-7938-3126
}
\fi
}

\maketitle

\begin{abstract}
This paper introduces \textit{RAGnaroX}, a resource-efficient ChatOps assistant that operates entirely on commodity hardware. Unlike existing solutions that often rely on external providers such as Azure or OpenAI, RAGnaroX offers a fully auditable, on-premise stack implemented in Rust. Its architecture integrates modular data ingestion, hybrid retrieval, and function calling, enabling flexible yet secure deployment. Our evaluation focuses on the RAG pipeline, with benchmarks conducted on the SQuAD (single-hop QA), MultiHopRAG (multi-hop QA), and MLQA (cross-lingual QA) datasets. Results show that RAGnaroX achieves competitive accuracy while maintaining strong resource efficiency, for example, reaching 0.90 context precision on single-hop questions with an average response time of 2.5 seconds per request. A replication package containing the tool, the demonstration video (\url{https://www.youtube.com/watch?v=cDxfuEbcoM4}), and all supporting materials are available at
\url{https://github.com/genius-itea/RAGnaroX.git}.
\end{abstract}


\section{Introduction}
By 2025, more than two-thirds of companies had integrated AI into their business operations across a variety of use cases \cite{rosenbushWhyCompaniesAre2025}. Thereby, a commonly seen technique is the enhancement of LLMs through knowledge integration, commonly referred to as Retrieval-Augmented Generation (RAG), which is increasingly combined with function-calling mechanisms. Consequently, as dependence on generative AI grows, so does the risk of vendor lock-in, particularly with major U.S.-based technology firms whose proprietary ecosystems dominate the market \cite{howcroftBanksSayGrowing2024}. At the same time, China is rapidly expanding its AI capabilities and investments, positioning itself as a significant competitor to U.S. dominance \cite{alshebliChinaUSProduce2024}. Overall, the tech industry is investing heavily in frontier LLMs, with compute demand projected to increase 2.25‑fold over the next two years \cite{kumarTrendsFrontierAI2025}. While these advancements are likely to enhance model quality, they are also expected to drive up usage costs. 

The combination of increasing external dependency and escalating costs poses a significant strategic risk to organizations. Additionally, in regulated fields such as medicine and finance, dependence on external AI providers also raises compliance concerns. External (proprietary) hosted models are often difficult to control or audit fully, exposing organizations to legal and reputational risks \cite{szadeczkyRiskRegulationGovernance2025}. 

Given the aforementioned challenges with external, commercial providers, we developed RAGnaroX, an on-premise, auditable RAG stack that combines the advantages of one of the most secure and performant programming languages, Rust \cite{bugdenRustProgrammingLanguage2022}, and llama.cpp \cite{ggmlLlamacpp} for quantized local inference of large and small language models (SLMs). We further integrated ChatOps, conversational agents embedded in operational workflows to directly execute system actions\cite{peciAgenticAIChatops2025}, into RAGnaroX via the Model Context Protocol (MCP) \cite{krishnanAdvancingMultiAgentSystems2025}, thereby rendering retrieved knowledge operational rather than merely informative. 

\section{Conception of RAGnaroX}

\begin{figure}
    \centering
    \includegraphics[width=\linewidth]{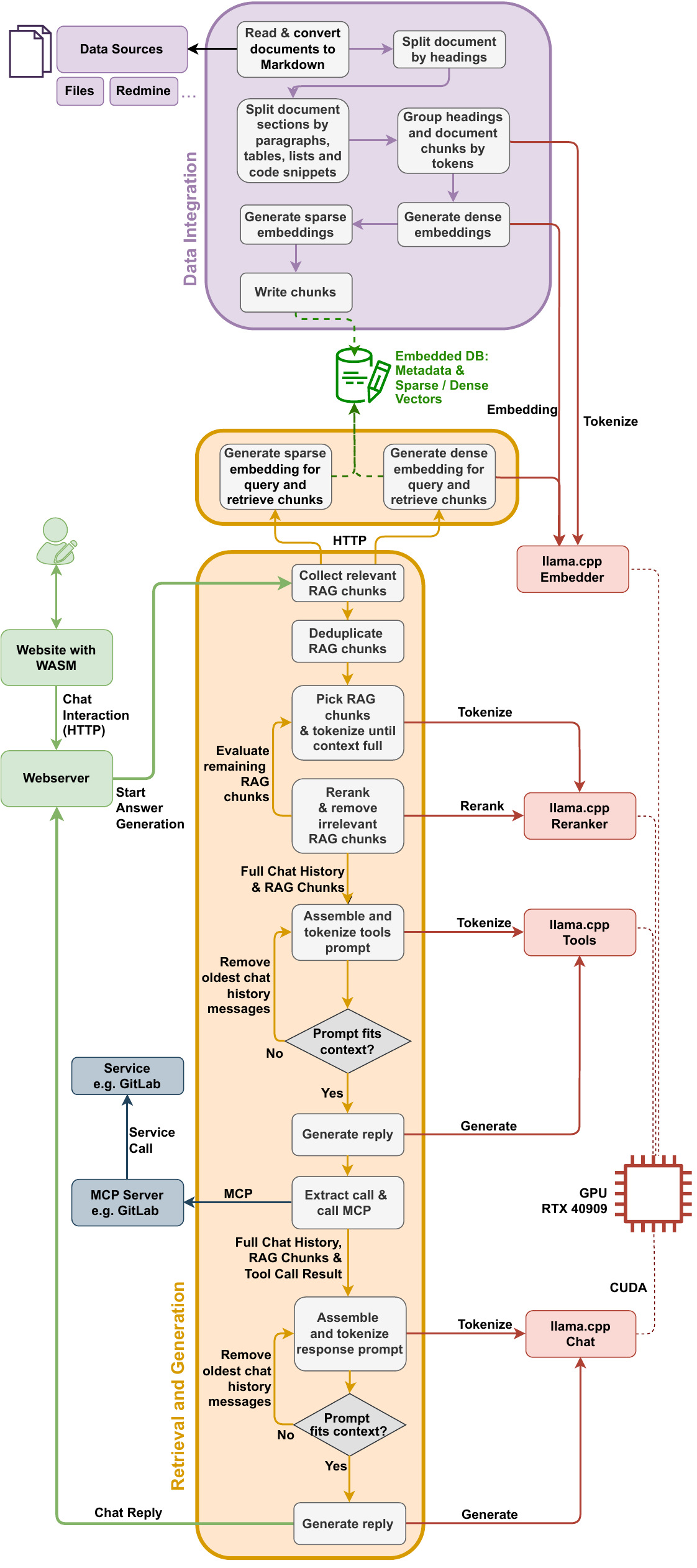}
    \caption{Conceptual overview of the RAGnaroX data integration and retrieval–generation pipeline for ChatOps.}
    \label{fig:architecture}
\end{figure}

To reduce integration and operational costs, the hardware requirements for running RAGnaroX were kept to a minimum. Therefore, the target configuration was a commodity computer equipped with 64~GB of RAM and a 24GB VRAM graphics card NVIDIA RTX 4090. However, we expect the requirements to continue declining as the quality of SLMs improves (e.g., RTX 4060). 

The interoperability with existing infrastructures and the adaptability to new requirements, both in software and hardware, were factors that led to the initial decision to adopt a Rust microservice architecture that utilizes HTTP and JSON for interprocess communication. As shown in \autoref{fig:architecture}, RAGnaroX’s conception is organized around two main components:

\paragraph{Data Integration Component} The modular adaptability enables the integration of various data sources (e.g., GitLab, Redmine) and data types (e.g., emails, files, wiki pages, issues).

In the first stage of the processing pipeline, the raw artifacts are converted to Markdown (e.g., from Textile), a format chosen because it seems to be highly LM interpretable, providing also simple syntactic markers and a rich toolset to transform to (e.g., PDF $\rightarrow$ Markdown) \cite{chenMDEvalEvaluatingEnhancing2025}. During the conversion phase, redundant repeating sequences (e.g., white spaces and dashes in table headers) are removed, thereby reducing the overall document size. This, in turn, decreases the storage footprint and optimizes the chunking process, enabling faster/efficient loading.

Each Markdown document is then divided by headings, preserving the document hierarchy. Each heading's contents are separated into paragraphs, tables, lists, and code snippets, an approach already positively supported by findings by Nguyen et al.\cite{guyenRAGHierarchicalTextSegmentation2025}. The resulting text blocks are paired with their headings and then tokenized to ensure that the text fits the context size of the embedding model (e.g., multilingual-e5-large-instruct). If the text is too large, it is divided by retaining the headings and headers for the tables, ensuring that each part of the information is placed in context, which is crucial for semantic search. Once the chunks are generated, their source and the timestamp of their generation are attached, and their dense and sparse vectors (BM25) are calculated and stored in \textit{Parquet} data format. 

\paragraph{Retrieval and Generation Component} Similar to the backend microservices, the RAGnaroX frontend is built in Rust. This enables the two to exchange data structures and logic. The frontend transmits the chat history to the backend, and the backend uses HTTP to reach all registered RAG source microservices and request all pertinent chunks. Every RAG source uses BM25 and semantic search (cosine similarity) to look for chunks in its embedded database. The hybrid approach to chunk retrieval is predicated on the complementary nature of both algorithms as well as their acknowledged advantages and disadvantages \cite{raoRetrieval2025}.

Duplicate entries are eliminated after every RAG source has returned its chunks. Next, a reranker SLM (e.g., bge-reranker-v2-m3) is used to reorder the remaining chunks and eliminate irrelevant information \cite{raoRetrieval2025}. Due to the finite context size of the reranker, the chunks are grouped by their size in tokens and iteratively reranked in batches until the desired number of chunks remains.

The next step, which is the foundation of ChatOps, involves using another specialized SLM (e.g., Qwen3-4B-Thinking-2507) to call registered functions via MCP. Due to constraints in the model context size, the prompt is generated iteratively. With each iteration, older chat history messages are removed until all chunks, the description of the available function calls, and the most recent messages in the chat history fit in the context. This procedure adapts to the different model context sizes and also preserves as much pertinent data as possible. Once the model decides which functions to call, the registered MCP endpoints (e.g., GitLab) are invoked, and their responses are added to the RAG chunks.

The final prompt to generate the user's answer is once again put together in the same manner as for function calling. Once the answer has been streamed token by token, the list of chunks is also transmitted to allow the user to verify the answer.

\subsection{Methodology for RAG Evaluation}
The evaluation of RAGnaroX centers on the information retrieval task, with three benchmark datasets selected to highlight different challenges commonly encountered in practical deployments:
\begin{enumerate}[label=(\roman*)]
  \item \textit{SQuAD v1.1}~\cite{rajpurkarSQuAD100000Questions2016}: a  single-hop factoid QA benchmark requiring answer extraction \textbf{from a single document}, 
  \item \textit{MultiHopRAG}~\cite{tangMultiHopRAGBenchmarkingRetrievalAugmented2024}: a benchmark for multi-hop reasoning, where answering a query requires integrating evidence across \textbf{multiple retrieved documents}, and 
  \item \textit{MLQA}~\cite{lewisMLQAEvaluatingCrosslingual2020}:  a cross-lingual QA benchmark for evaluating transfer and retrieval \textbf{across languages}. In these experiments, we focused on English, Spanish, and German.
\end{enumerate}

To run the RAG experiments, we used an RTX 4090 GPU. If not otherwise mentioned, we employed \textit{Qwen3-4B-q8} for generation, \textit{multilingual-e5-large-q8} for embedding, and \textit{bge-reranker-v2-m3-q8} for reranking, with 350 chunking tokens, and 3 chunks (given as \texttt{@3}). Different components were modified to make them comparable to existing benchmarks \cite{tangMultiHopRAGBenchmarkingRetrievalAugmented2024, sawarkarBlendedRAGImproving2024}.

We conducted our evaluation with RAGAS~\cite{esRAGAsAutomatedEvaluation2024}. Prior tests with \textit{gpt-oss-20B}, \textit{Qwen3-30B}, and \textit{gpt-oss-120B} showed that results were largely invariant to evaluator size, with deviations $\leq 0.02$ for retrieval metrics (\textit{Context Recall}@3, \textit{Context Precision}@3) and $\leq 0.03$ for generation metrics  (\textit{Faithfulness@3}, \textit{Answer Relevancy@3}) on SQuAD. We selected gpt-oss-20B for its efficiency and to reduce potential bias toward the models under evaluation.

Apart from that, for each test-run, we logged execution traces using \texttt{strace} for system calls and \texttt{cProfile} for function-level profiling. GPU activity was sampled via \texttt{nvidia-smi}, while CPU and RAM usage were aligned with \texttt{/proc} metrics.

\section{Results and Discussion}

\begin{figure*}[t]
    \centering
    \includegraphics[width=\linewidth]{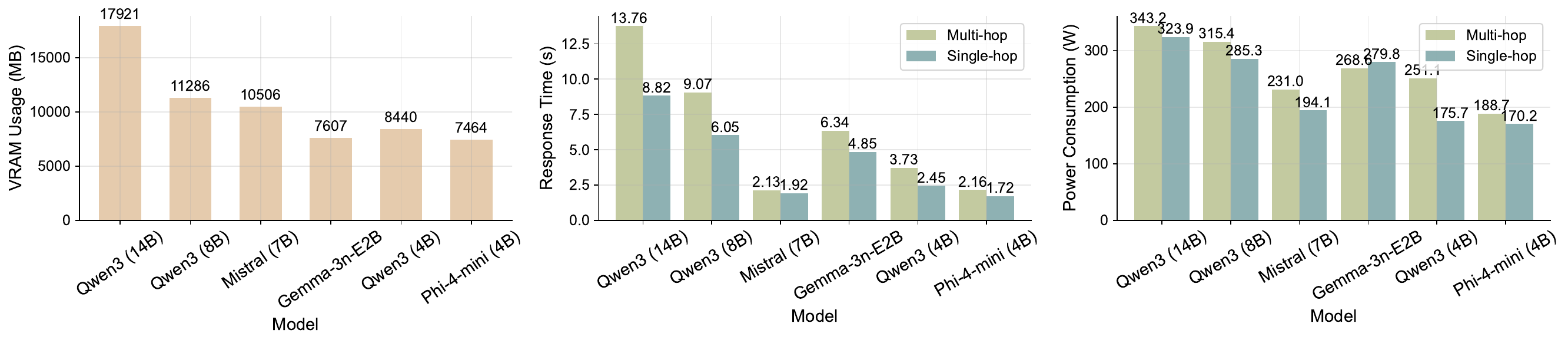}
    \caption{Performance Metrics in RAGnaroX, with changed generation models}
    \label{fig:rag-performance}
\end{figure*}

\subsubsection{Single-Hop-Dataset (SQuAD v1.1)} The results indicate that the retrieval achieved high accuracy, with an average \textit{Context Precision@5} of $0.90$ and \textit{Context Recall@5} of $0.94$. In comparison to the Blended-RAG implementation by Sawarkar et al. \cite{sawarkarBlendedRAGImproving2024}, which also assessed retrieval quality using the top-5 documents, our retrieval component demonstrates performance on par with their different RAG approaches, yielding a comparable level of effectiveness (min.: 90.7\% – max.: 94.89\%). 

\begin{table}[!t]
\caption{Faithfulness and answer relevancy comparison}
\label{tab:faith-ansrel}
\centering
\begin{tabular}{lcccc}
\toprule
\multirow{2}{*}{Model} &
\multicolumn{2}{c}{Single-hop QA} &
\multicolumn{2}{c}{Multi-hop QA} \\
\cmidrule(lr){2-3} \cmidrule(lr){4-5}
 & Faith. & AnsRel. & Faith. & AnsRel. \\
\midrule
Qwen3 (14B)       & 0.8327 & 0.7846 & \textbf{0.7039} & 0.6407 \\
Qwen3 (8B)        & 0.8364 & 0.7865 & 0.6874 & 0.6328 \\
Mistral (7B)      & 0.7963 & 0.7705 & 0.4654 & 0.6568 \\
Gemma-3n-E2B      & 0.7183 & 0.7722 & 0.4067 & 0.5056 \\
Qwen3 (4B)        & \textbf{0.8588} & \textbf{0.8168} & 0.6341 & \textbf{0.6790} \\
Phi-4-mini (4B)   & 0.7192 & 0.7764 & 0.4583 & 0.6612 \\
\bottomrule
\end{tabular}
\end{table}

Using the retrieved chunks, we evaluated five SLMs for answer generation, including Qwen3 models of varying sizes, as well as other recent baselines, as presented in  \autoref{tab:faith-ansrel}. Notably, the smallest model, Qwen3 (4B), illustrates that increasing model size does not necessarily translate into improved grounding or factual consistency. Indeed, its more limited reliance on parametric knowledge may even constitute an advantage in contexts of knowledge conflict, when retrieved evidence diverges from the model’s internal representations, an issue that approaches such as Zhang et al. \cite{zhangFaithfulRAGFactLevelConflict2025} explicitly aim to address.

\subsubsection{Multi-Hop-Dataset (MultiHopRAG)} In the multi-hop setting, retrieval quality declines, with \textit{Context Precision@4} = 0.42 and \textit{Context Recall@4} = 0.52. Compared to the benchmark metrics reported by Tang et al.~\cite{tangMultiHopRAGBenchmarkingRetrievalAugmented2024}, RAGnaroX attains \textit{Hits@4} = 0.57, following their evaluation strategy. This places RAGnaroX's retrieval on par with the other multilingual embedding models, such as \textit{intfloat/e5-base-v2} and \textit{hkunlp/instructor-large}, but still behind English embedders, e.g. \textit{bge-large-en-v1.5} or \textit{text-embedding-ada-002}. Thus, further enhancements are needed, such as a knowledge graph-based inclusion, which might further enhance the multi-hop retrieval \cite{hanRAGVsGraphRAG2025}. Compared to single-hop results, Qwen3 (4B) continues to achieve the highest answer relevancy. In contrast, faithfulness benefits from SLMs with more parameters, likely due to a decrease of the number of retrieved documents and the consequent reliance on the SLM’s internal knowledge.

\begin{table}[!t]
\caption{RAG system performance of Qwen3 14B and 4B}
\label{tab:rag-performance}
\centering
\begin{tabular}{lcccccc}
\toprule
\multirow{2}{*}{\makecell{Corpus--\\Question}} &
\multirow{2}{*}{Ctx.\ P.} &
\multirow{2}{*}{Ctx.\ R.} &
\multicolumn{2}{c}{Faith.} &
\multicolumn{2}{c}{AnsRel.} \\
\cmidrule(lr){4-5} \cmidrule(lr){6-7}
 &  &  & 14B & 4B & 14B & 4B \\
\midrule
en--en & 0.86 & 0.91 & \textbf{0.83} & 0.81 & 0.73 & \textbf{0.75} \\
de--de & 0.74 & 0.77 & \textbf{0.77} & 0.73 & 0.41 & \textbf{0.44} \\
es--es & 0.82 & 0.87 & 0.77 & \textbf{0.80} & 0.53 & \textbf{0.58} \\
en--de & 0.59 & 0.70 & \textbf{0.74} & 0.70 & 0.41 & \textbf{0.43} \\
de--en & 0.64 & 0.73 & \textbf{0.73} & 0.71 & \textbf{0.71} & 0.65 \\
en--es & 0.64 & 0.71 & \textbf{0.74} & 0.73 & \textbf{0.54} & 0.52 \\
de--es & 0.50 & 0.63 & \textbf{0.71} & 0.61 & \textbf{0.52} & 0.47 \\
es--en & 0.71 & 0.81 & \textbf{0.75} & 0.73 & 0.69 & \textbf{0.69} \\
es--de & 0.52 & 0.72 & \textbf{0.70} & 0.65 & \textbf{0.45} & 0.44 \\
\bottomrule
\end{tabular}
\end{table}

\subsubsection{Multi-Language-Dataset (MLQA)} Having a look at different language configurations, given in Table~\ref{tab:rag-performance}, it is derivable that en-en performs pretty well, as model weights might mainly be trained on the English corpus, specifically for the retrieval part. If we consider non-English languages with the same data corpus and questions (e.g., de-de), the performance drops; if the corpus and language differ, the performance drops even further. It can be seen that here a larger model Qwen3-14B clearly outperforms smaller models Qwen3-4b.

\subsubsection{Energy-Drawn} Since the SLMs are executed locally, see \autoref{fig:rag-performance}, energy consumption must also be considered. In this regard, \textit{Phi-4-mini} performs best. Interestingly, \textit{Qwen} exhibits significantly higher energy consumption in multi-hop scenarios, which may be attributed to longer reasoning times. 

\subsubsection{Response Latency} 
An essential aspect of a ChatOps assistant is its performance, particularly its responsiveness. The results, illustrated in \autoref{fig:rag-performance}, indicate that pipelines employing generative models with fewer than 5B parameters achieve response times below 2.5~s for single-hop queries and 3.8~s for multi-hop queries. Such delays are considered satisfactory from a user perspective, based on the findings by Maslych et al \cite{maslychMitigatingResponseDelays2025}.

\begin{mybox}[title=Comparing RAGnaroX in Real-World QA] Beyond the benchmark results, we further evaluated our approach on a documentation-based use case from \textit{\ifanonymized HotelSoftwareCompany\else  CASBLANCA hotelsoftware \fi }, employing 250 factoid QA pairs. In this setting, \ifanonymized HotelSoftwareCompany\else  CASBLANCA  \fi’s production-ready RAG system, built with an agentic approach leveraging Azure components and the latest commercial models, achieved 7\% higher Context Recall@10 (RAGnaroX: 0.83, \ifanonymized HotelSoftwareCompany \else  CASABLANCA  \fi RAG: 0.90), while maintaining nearly the same answer relevancy against RAGnaroX. For generation metrics, RAGnaroX exhibited 12\% lower Faithfulness@10 when \ifanonymized HotelSoftwareCompany \else  CASBLANCA \fi `s RAG used GPT-4.1 for generation, but comparable results when they used GPT-4.1-mini. This comparison suggests RAGnaroX is capable of handling real-world data. Nevertheless, further evaluation and empirical evidence are required to compare its effectiveness in real-world settings (performance, costs, usability, etc.).
\end{mybox}

\section{Conclusion and Future Outlook}
RAGnaroX is a resource-efficient architecture that can serve as the foundation for various use cases, featuring a function-calling mechanism and secure on-site operability. A concrete example can be seen in the demonstration video, where a ticket support system is simulated: A customer support representative can access the documentation information of \ifanonymized HotelSoftwareCompany \else  CASBLANCA hotelsoftware \fi , and create new issues via chat. Overall, focusing on the information retrieval part, the resource-efficient conception performs reliably for single-faceted questions within acceptable response times. Based on the findings, we will focus on two directions: (1) extending support for multi-hop questions through knowledge-graph integration, and (2) trying to improve cross-language performance through pre-translation. Furthermore, future work will include benchmarks for function calling. 

\bibliographystyle{IEEEtran}
\bibliography{references}

\end{document}